\documentclass[twocolumn,longbibliography]{revtex4-1}
\usepackage{xparse}
\NewDocumentCommand{\dslash}{s}{%
  \IfBooleanTF{#1}
    {\big/\mkern-7mu\big/}
    {/\mkern-6mu/}%
}
\usepackage{graphicx}
\usepackage{amsmath,amssymb}
\usepackage{epstopdf}
\usepackage{tikz}
\usepackage{mathptmx} 

\newcommand{\be}{\begin{equation}}
\newcommand{\ee}{\end{equation}}

\usepackage{hyperref}
\DeclareMathAlphabet{\mathcal}{OMS}{cmsy}{m}{n}

\begin{document}
\title{Three-dimensional topological photonic crystal with a single surface Dirac cone}
\author{Ling Lu$^{*}$$^{\dagger}$$^{1}$, Chen Fang$^{\dagger}$$^1$, Liang Fu$^{*}$$^1$, Steven G. Johnson$^2$, John D. Joannopoulos$^1$ and Marin Solja\v{c}i\'{c}$^1$}
\address{
$^{1}$Department of Physics, $^{2}$Department of Mathematics, Massachusetts Institute of Technology, Cambridge, Massachusetts}
\vspace{-0.2in}
\begin{abstract}
A single Dirac cone on the surface is the hallmark of three-dimensional~(3D) topological insulators, where the double degeneracy at the Dirac point is protected by time-reversal symmetry and the spin-splitting away from the point is provided by the spin-orbital coupling.
Here we predict a single Dirac-cone surface state in a 3D photonic crystal, where the degeneracy at the Dirac point is protected by a nonsymmorphic glide reflection and the linear splitting away from it is enabled by breaking time-reversal symmetry.
Such a gapless surface state is fully robust against random disorder of any type.
This bosonic topological band structure is achieved by applying alternating magnetization to gap out the 3D ``generalized
Dirac points'' discovered in the bulk of our crystal.
The $Z_2$ bulk invariant is characterized through the evolution of Wannier centers.
Our proposal---readily realizable using ferrimagnetic materials at microwave frequencies---can also be regarded as the photonic analog of topological crystalline insulators, providing the first 3D bosonic symmetry-protected topological system.
\end{abstract}
\email{linglu@mit.edu; liangfu@mit.edu\\
$^{\dagger}$The first two authors contributed equally to this work.}


\maketitle

Topology of electron wavefunctions was first introduced to characterize the quantum Hall states in two dimensions~(2D) discovered in 1980~\cite{Thouless1982}.
Over the past decade, it has been recognized that symmetry plays a crucial role in the classification of topological phases, 
leading to the broad notion of symmetry-protected topological phases~\cite{Schnyder2008,Chen2012,Wang2014}.
As a primary example, topological insulators~\cite{Hasan2010,Qi2011,Moore2010birth} are distinguished from normal insulators in the presence of time-reversal symmetry~($\mathcal{T}$).
A 3D topological insulator exhibits an odd number of protected surface Dirac cones, a unique property that cannot be realized in any 2D systems.
Importantly, the existence of topological insulators requires Kramers' degeneracy in spin-orbit coupled electronic materials; this forbids any direct analogue in boson systems~\cite{lu2014topological}. 
In this report, we discover a 3D topological photonic crystal phase hosting a single surface Dirac cone, which is protected by a crystal symmetry~\cite{Fu2011,ando2015topological,chiu2015classification,fang2015new} --- the nonsymmorphic glide reflection
rather than $\mathcal{T}$. Our finding expands the scope of 3D topological materials from fermions to bosons.

Unlike in Fermi systems, achieving a single Dirac cone in boson systems requires $\mathcal{T}$ breaking. This is because the $T$ operator acts differently on bosons and fermions: $\mathcal{T}_f^2=-1$ for fermions with half-integer spins and $\mathcal{T}_b^2=1$ for bosons with integer spins. As a result, $\mathcal{T}_b$ is not compatible with the Hamiltonian of a single Dirac cone, while $\mathcal{T}_f$ is (see Supplementary Information).
Instead of $\mathcal{T}_f$, the Dirac point-degeneracy in our photonic crystal is protected by a glide reflection~\cite{liu2014topological,fang2015new,shiozaki2015z} , which ensures a band-crossing on high-symmetry lines.
This crystal-symmetry-protected topological photonic crystal can be regarded as a bosonic analog of the recently discovered topological crystalline insulators in electronic systems~\cite{Fu2011,Hsieh2012,Dziawa2012,Tanaka2012,Xu2012}. 

\begin{figure*}[!ht]
\centering
\includegraphics[width=\textwidth]{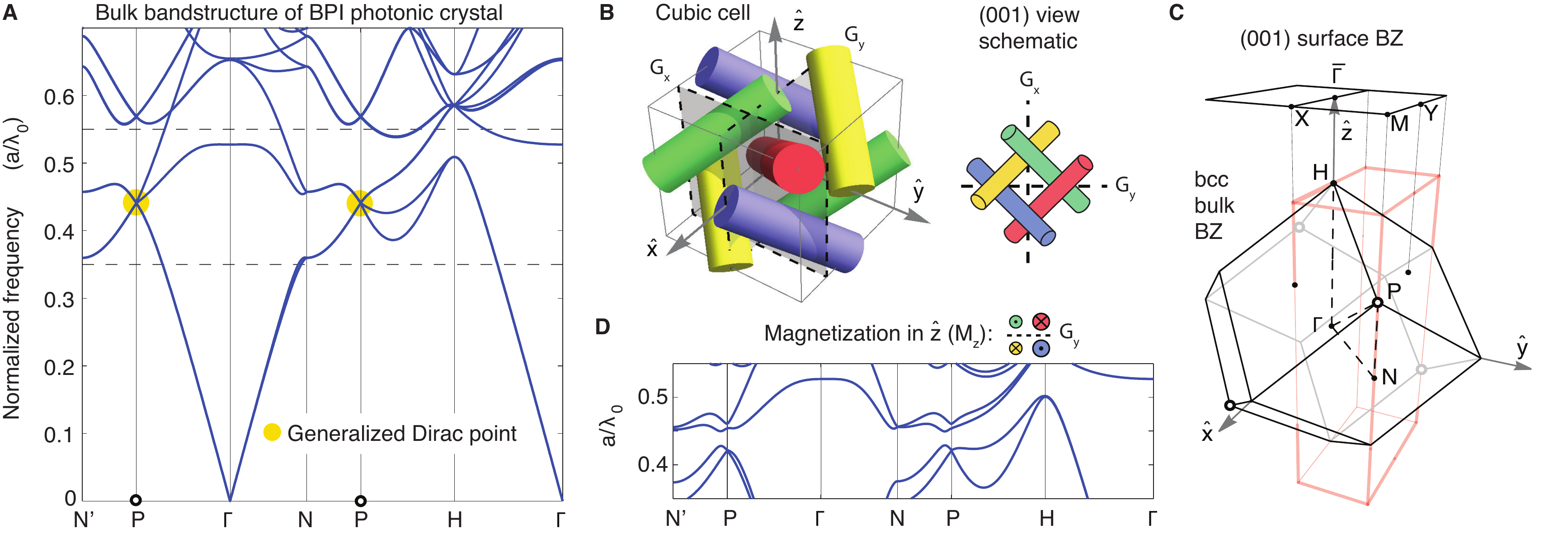}
\caption{Bulk band structures of the BPI photonic crystal.
A) The GDP is between the first four bands.
B) The cubic unit cell of length $a$ consisting of four identical dielectric rods oriented along the bcc lattice vectors of $(111)$~(red), $(\bar{1}11)$~(yellow), $(1\bar{1}1)$~(blue) and $(11\bar{1})$~(green). The rods go through (0,0,0)$a$, (0,0.5,0)$a$. (0.5,0,0)$a$ and (0,0,0.5)$a$ respectively.
There are two glide reflection planes~($G_x$ and $G_y$) in the structure, invariant on the (001) surface.
The top-view schematic illustrates the relations between the rods under operations of $G_x$ and $G_y$.
C) The bcc BZ and its projection onto the (001) surface BZ. The transparent red box outlines the volume in the bulk BZ that projects to half of the surface BZ.
D) Bulk band structure showing that the GDP opens when magnetization is applied on the rods without breaking $G_y$.
}
\label{Fig:1}
\end{figure*}

Our starting point is a photonic crystal having a body-centered-cubic~(bcc) unit cell which contains four identical dielectric rods, illustrated with different colors for clarity in Fig. \ref{Fig:1}B.
This crystal belongs to the nonsymmorphic space group of No. 230~($Ia\bar{3}d$) that contains glide reflections and inversion.
Interestingly, such a triply-periodic structure self-assembles as disclination-line networks in the first blue phase of liquid crystals~\cite{meiboom1983lattice}, denoted as BPI.
Here, the dielectric constant~($\epsilon$) of the rods is 11 and radius is $0.13a$ where $a$ is the length of the cubic cell.
In Fig. \ref{Fig:1}A , the photonic band structure of BPI shows a four-fold degenerate point at the $P$ momentum, dispersing linearly in all directions of the 3D momentum space.
Unlike a 3D Dirac point~\cite{Young20123D-Dirac,liu2014discovery} --- a four-fold degeneracy point which splits into two sets of doublet bands along any direction --- our four-fold degeneracy here splits into four bands along a generic direction. 
In Fig. \ref{Fig:1}A, this splitting is not obvious, since most dispersions still remain doubly-degenerate along high-symmetry momentum lines.
However, it is clear that the 3rd and 4th bands split along $P-\Gamma$ and the 1st and 2nd bands split along $P-H$.
We name this type of degeneracy a 3D ``generalized Dirac point''~(GDP).
We note that there are two non-equivalent $P$ points in the 3D bcc Brillouin zone~(BZ) related by inversion.
Interestingly, the two GDPs~(at $\pm{P}$) are the only band-touchings between band 1,2 and 3,4.
When the space group is perturbed, the GDPs could turn into line nodes, Weyl points~\cite{lu2013weyl,lu2015experimental} or open bandgaps.
Detailed studies of GDPs will be presented in another paper.


In symmorphic space groups, where the point groups decouple from lattice translations, the highest dimension of group representation is three. The four-fold band degeneracies of the GDPs are hence the consequence of the non-symmorphic symmetries of glide reflections and screw rotations in BPI.
A non-symmorphic symmetry is in general composed of a point group~(mirror or rotation) followed by a fractional lattice translation, where neither of the two is a symmetry of the system. 
The important feature of a non-symmorphic space group is the extra band degeneracies at the BZ boundaries~\cite{mock2010space,lu2012three,parameswaran2013topological,young2015dirac}. Since the screw rotations cannot be preserved on a planar surface, we focus on the glide reflections in order to obtain protected surface states.
Shown in Fig. \ref{Fig:1}B, the (001) surface has two glide reflections $G_x=\{M_x|\frac{a}{2}\hat{x}-\frac{a}{2}\hat{y}\}$ and $G_y=\{M_y|\frac{a}{2}\hat{x}\}$. 
The top view schematic illustrates the relations between the four rods under the two glide reflections.
The (001) surface BZ is plotted in Fig. \ref{Fig:1}C.

\begin{figure}[!h]
\centering
\includegraphics[width=0.5\textwidth]{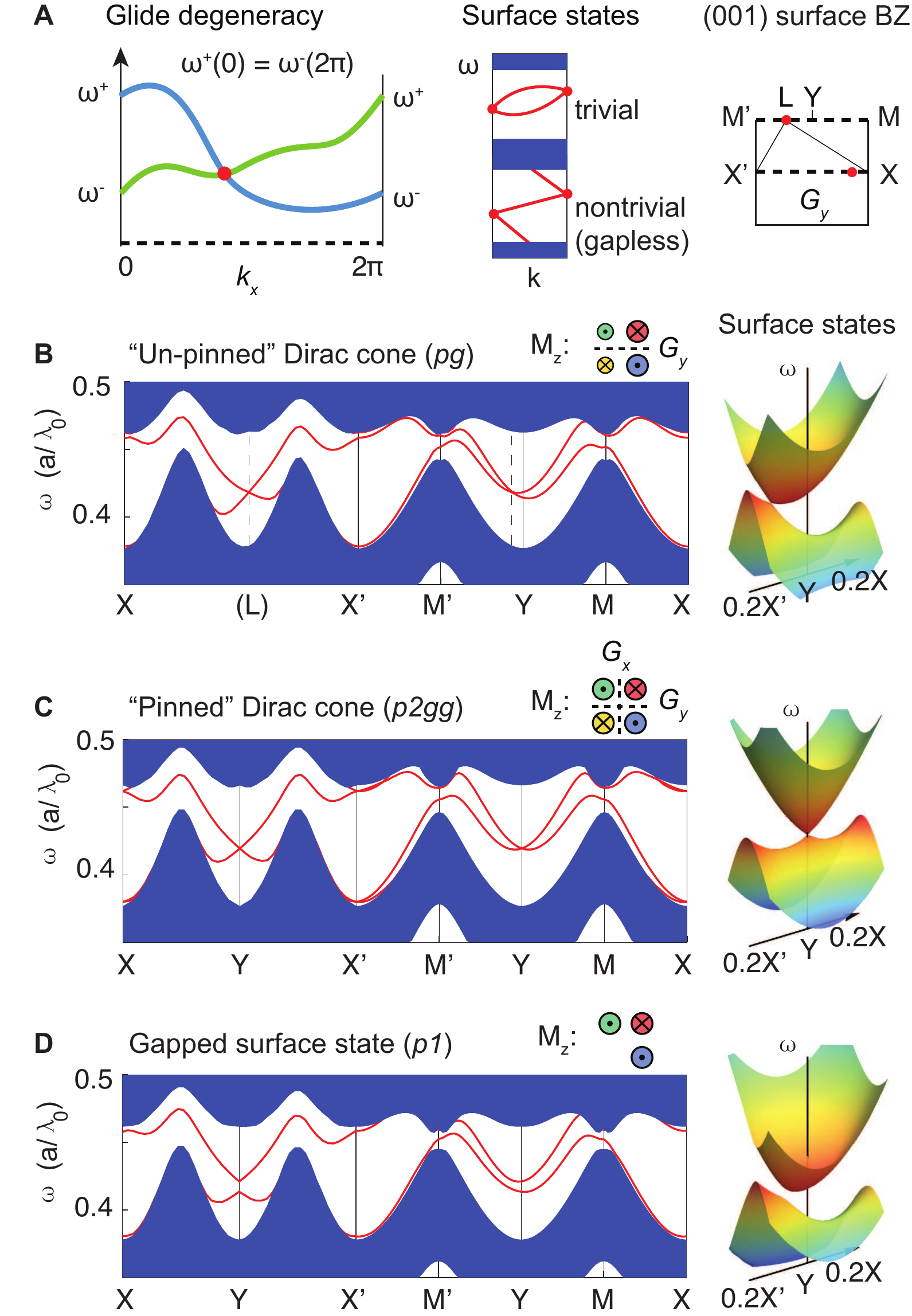}
\caption{The (001) surface states after breaking $\mathcal{T}$. The surface and bulk dispersions are plotted in red and blue colors, respectively.
A) Illustration of the two crossing points~(double-degeneracies) on the $G_y$-invariant lines of $M'-M$ and $X'-X$~(dotted) in the surface BZ.
Consequently, the surface states can have a gapless connectivity which is topologically nontrivial.
$M'$ and $M$ are the same point in the BZ, so are $X'$ and $X$.
B) Single Dirac cone at $L$ movable along the $M'-M$ line protected by $G_y$.
C) Single Dirac cone in B) pinned at $Y$ point due to the co-existence of $G_x$ and $G_y$.
D) Surface states in B) and C) gapped by breaking the glide-reflection symmetries.
For all above surface calculations, a perfect metallic boundary is placed on top, at the center of the cubic cell, on the (001) surface with an air gap spacing of 0.5$a$ from the photonic crystal surface. The 3D plots of the surface dispersions are plotted on the right to show the behavior of the surface Dirac cone. The 3D surface plots are centered at the $Y$ point, with a span of $0.2\pi/a$ in $k_x$ and $0.1\pi/a$ in $k_y$ and a normalized frequency range between 0.41 and 0.43.
}
\label{Fig:2}
\end{figure}

A glide reflection ensures a linear point-degeneracy along each glide-reflection-invariant momentum line.
To see this, we study the Bloch states on the $G_y$-invariant lines of $X'-X$ and $M'-M$ dashed in the (001) surface BZ on the right of Fig. \ref{Fig:2}A.
A Bloch state with momentum $(k_x,k_y)$ is mapped to another state with momentum $(k_x,-k_y)$ under $G_y$, so for any state along these two lines with $k_y=0$ and $k_y=\pi/a$, its momentum is invariant under $G_y$. This means the eigenvalues of $G_y$ ($g_y$) are good quantum numbers for the Bloch states on these two lines.
Since ${G_y}^2=\{1|a\hat{x}\}$, ${g_y(k_x)}=\pm e^{-i{k_x}\frac{a}{2}}$ [${g_y^2(k_x)}=e^{-i{k_x}a}$] which is  $k_x$-dependent.
The two branches of glide-reflection eigenvalues always differ by a minus sign and they evolve into each other after a 2$\pi$ transportation along the $G_y$-invariant lines due to the fact that $g_y(k)=-g_y(k+\frac{2\pi}{a})$.
As a result, the corresponding wavefunctions of the two branches have the same winding as their eigenvalues---a unique property of the half-lattice translation in glide reflections.
Consequently, the two frequency eigenvalues of the two Bloch modes also switch values after transporting a period along the invariant momentum lines, illustrated in Fig. \ref{Fig:2}A.
Assume the frequencies of the two modes are $\omega^+$ and $\omega^-$ at an arbitrary $k_x$ point~(say $k_xa=0$). The frequency dispersions switch their values at $ka=2\pi$.
This switch ensures a crossing point~(red dot) on $X'-X$ and $M'-M$ respectively.
We argue that these two protected double-degeneracies give a $Z_2$ classification of the surface states~\cite{fang2015new}.
Illustrated in the middle of \ref{Fig:2}A, there are two topologically in-equivalent ways for these two point-degeneracies to connect. The gapless connection is a signature of the topologically nontrivial surface states protected by $G_y$.

We now break $\mathcal{T}$ in the BPI photonic crystal to open the bulk bandgap without breaking the $G_y$.
Shown in Fig. \ref{Fig:1}D, the GDP at $P$ point lift up into a bandgap when we apply alternating magnetization on the rods along $\hat{z}$.
These magnetization induces off-diagonal imaginary parts in the dielectric constant~($\epsilon$) of materials with gyroelectric response~\cite{Haldane:2008-PRL}. (Ferrimagnetic materials of gyromagnetic response~\cite{Wang2009} give the same results in Supplementary Information). Here $\mu=1$ and $\epsilon=\left(
\begin{array}{ccc}
\epsilon_{\dslash} & \kappa & 0\\
-\kappa & \epsilon_{\dslash} &  0\\
0 &0 & \epsilon_{zz}\\
\end{array}
\right)$, where $\epsilon_{zz}=11$, $\epsilon_{\dslash}^2-|\kappa|^2=\epsilon_{zz}^2$~\cite{lu2013weyl} and $\kappa$ is a non-zero imaginary number when the magnetization~($M_z$) is present.
In Fig. \ref{Fig:1}D, $\kappa=-10i,-5i,+5i,+10i$ for the red, yellow, blue and green rods respectively.
This configuration preserves $G_y$, because magnetization~(magnetic field) flips sign under a mirror~(reflection) operation.
The 2D plane group of the resulting (001) surface is $pg$.

The (001) surface state, plotted in Fig. \ref{Fig:2}B, has a single Dirac cone at point $L$ on the $M'-M$ line, consistent with the glide-reflection degeneracy in Fig. \ref{Fig:2}A.
By varying the magnetization or rod radius without breaking $G_y$, the Dirac point $L$ moves along the $G_y$ invariant line $M'-M$.
This single Dirac cone at $L$ is connected gaplessly with the bulk bands across the band gap.
In Fig. \ref{Fig:2}C, we restore $G_x$ to coexist with $G_y$ by doubling the magnetization amplitude of the green and yellow rods~($|\kappa|$ from 5 to 10).
The surface plane group becomes $p2gg$. Due to this extra glide-reflection plane through $Y$ point, the surface Dirac cone is then pinned at $Y$ on $M'-M$.
If we break both glide symmetries by de-magnetizing the yellow rod, both glide planes of $G_x$ and $G_y$ are broken and the surface plane group reduces to $p1$.
The surface Dirac cone is now gapped as shown in Fig. \ref{Fig:2}D. This demonstrates that the gapless surface states are indeed protected by the glide reflection.

\begin{figure}[h!]
\centering
\includegraphics[width=0.5\textwidth]{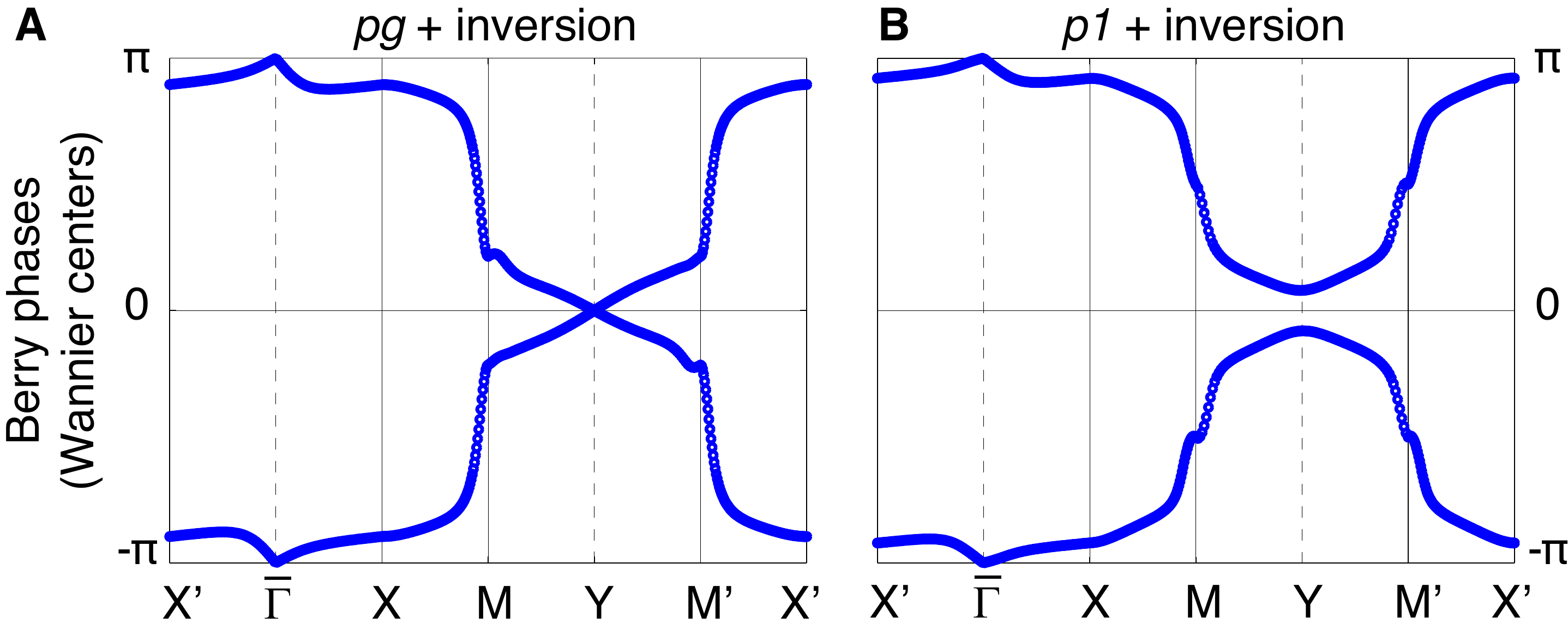}
\caption{Hybrid Wannier centers in the surface BZ indicating bulk topologies and the connections of the surface states.
(A) The gapless hybrid Wannier centers corresponds to the non-trivial surface states~($pg$) in Fig. \ref{Fig:2}B.
(B) The gapped bybrid Wannier centers corresponds to the trivial surface states~($p1$) in Fig. \ref{Fig:2}D.
The hybrid Wannier centers corresponds to the $p2gg$ surface in Fig. \ref{Fig:2}C is plotted in the Supplementary Information.
}
\label{Fig:3}
\end{figure}
The principle of bulk-edge correspondence 
says that the surface state is a holographic representation of the bulk topology. We demonstrate this correspondence between the surface states in Fig. \ref{Fig:2} and the ``hybrid Wannier centers"~\cite{taherinejad2014wannier} of the bulk bands below the bandgap computed in Fig. \ref{Fig:3}. This approach is also known as the Wilson loops~\cite{yu2011equivalent,alexandradinata2014wilson}.
The hybrid Wannier function of each band is a spatially-localized wavefunction along $z$, obtained from Fourier-transforming the Bloch wavefunctions with respect to $k_z$ while keeping the other two surface momenta.
The $z$-position expectation values of the hybrid Wannier wavefuntions, i.e., the hybrid Wannier centers, are equivalent to the Berry phases of the bulk bands below the gap along a loop in $\hat{z}$ in the bulk BZ.
In our bcc lattice, this non-contractable loop (of length $4\pi/a$) is the vector connecting $H$ and $-H$ in Fig. \ref{Fig:1}C. This hybrid Wannier center is well defined up to a lattice period of $a/2$ in $\hat{z}$, and similarly, the Berry phase has a $2\pi$ phase ambiguity.
The calculations of the Berry phases are detailed in the Supplementary Information.

In the surface BZ, a gapless spectrum of Wannier centers~(Berry phases) indicates a non-trivial bulk topology and a gapless surface state.
In contrast, a gapped spectrum represents a trivial bulk topology and the absence of gapless surface states.
This can be understood by the following intuitive arguments. If there is a full gap in the spectrum of Wannier centers, then there is a certain position in $z$ where no state is localized. Terminating the bulk at that plane results in a surface without surface states --- the trivial surface states. On the other hand, if the Wannier center plot is gapless, then for any surface termination there must be a localized surface state at some surface momentum. The surface is hence gapless for terminations at arbitrary $z$ --- a nontrivial surface.
In Fig. \ref{Fig:3}, we plot the Wannier centers of the two lowest bands along the closed loop of $X'-X-M-M'-X'$ in the surface BZ.
Fig. \ref{Fig:3}A depicts the hybrid Wannier centers calculated for the bulk bands in Fig. \ref{Fig:1}D, whose surface state is shown in Fig. \ref{Fig:2}B. Similarly, the hybrid Wannier centers in Fig. \ref{Fig:3}B correspond to the surface states shown in Fig. \ref{Fig:2}D.
The Wannier centers are gapless in Fig. \ref{Fig:3}A, consistent with the existence of the gapless single surface Dirac cone in Fig. \ref{Fig:2}B.
In comparison, the Wannier centers in Fig. \ref{Fig:3}B is gapped, also consistent with the absence of topological surface states in Fig. \ref{Fig:2}D.
These data confirm the bulk-edge correspondence that the Wannier centers for all bulk bands below the bandgap is homotopic to the surface band structure of a semi-infinite system with one open surface.

Single-Dirac-cone surface states are fully robust and do not localize under arbitrary random disorders on the surface.
This has been discussed in 3D topological insulators where the surface states remain delocalized under random impurities of any type	~\cite{Fu2012,Fulga2014}.
In our case, although individual defects break the glide reflection, their ensemble average do not. Intuitively, if one local disorder generates a positive Dirac mass term within a region on the surface, there must be a neighboring region where the mass term is negative. A chiral edge mode exists along the edge between two regions with opposite masses, similar to the photonic one-way edge states~\cite{Haldane:2008-PRL,Wang2009} analogous to the quantum Hall effect. In the presence of strong disorder, these chiral edge modes percolate the surface and the surface states remain delocalized. The surface with a strong random disorder can be mapped to the electronic states at the critical point of a quantum Hall plateau transition, where chiral edge modes between regions of different Landau-level filling factors percolate. The transmission rate of light on the surface hence exhibits the universal scaling laws in the universality class of the quantum Hall plateau transitions~\cite{fang2015new,Ludwig1994}. Free from any interaction, this single-Dirac-cone surface state is an ideal platform for studying the critical phenomena of ``metal-insulator" transitions in Dirac systems~\cite{Bardarson2007,Fu2012}.

In 2D photonic crystals, topological band structrues protected by $\epsilon-\mu$ symmetry~\cite{khanikaev2013photonic,chen2014experimental} have been studied. However, symmetries in constitutive relations are difficult to maintain over a wide frequency bandwidth.
Another 2D example discusses the bulk topology of $C_6$ rotation~\cite{wu2015scheme}. Unfortunately, six-fold rotation cannot be preserved on the 1D edge and cannot protect edge states.
In contrast, our glide reflection can be maintained for all materials at all frequencies with protected surface states.


Experimentally, the $\mathcal{T}$-breaking BPI photonic crystals can be readily realized at microwave frequencies by assembling ferrimagnetic rods~\cite{Wang2009,skirlo2015experimental} with internal remnant magnetization without the need for external magnetic fields. These materials are commercially available, such as yttrium iron garnet.
Towards optical frequencies, $\mathcal{T}$-breaking could potentially be implemented through dynamic Floquet modulations~\cite{fang2012realizing,rechtsman2013photonic}.
In addition, our approach for photons can directly be used for phonons where $\mathcal{T}$-breaking can be achieved by spinning the rods~\cite{wang2015topological}.

This work demonstrates that symmetry-protected 3D topological bandgaps supporting disorder-immune surface states can be obtained in bosonic systems.
Spatial symmetries~\cite{Fu2011,chiu2015classification,liu2013antiferromagnetic,alexandradinata2014spin,fang2015new} other than the glide reflection are to be studied in the rich context of 230 space groups and 1651 magnetic groups for any bosonic particles.

We thank Timothy H. Hsieh, Aris Alexandradinata, B. Andrei Bernevig, Scott Skirlo, Abby Men, Junwei Liu, Fan Wang for discussions.
S.J. and J.J. were supported in part by the U.S.A.R.O. through the ISN, under Contract No. W911NF-13-D-0001.
C.F. and L.F. were supported by the DOE Office of Basic Energy Sciences, Division of Materials Sciences and Engineering under Award No. DE-SC0010526.
L.L. was supported in part by the MRSEC Program of the NSF under Award No. DMR-1419807.
M.S. and L.L. (analysis and reading of the manuscript) were supported in part by the MIT S3TEC EFRC of DOE under Grant No. DE-SC0001299.
\bibliography{Ling}

\clearpage
\makeatletter
\setcounter{figure}{0}
\renewcommand{\thefigure}{S\@arabic\c@figure}
\makeatother
\section*{Supplementary Information}
\subsection{Compatibility between time-reversal symmetries and the single-Dirac-cone surface state}
The key ingredient in achieving bosonic single Dirac surface states is the breaking of $\mathcal{T}$, which we prove by contradiction.
The $\mathcal{T}$ operator differs fundamentally for particles with different spins : $\mathcal{T}_f^2=-1$ for fermions with half-integer spins while $\mathcal{T}_b^2=1$ for bosons with integer spins.
Up to a choice of basis, the anti-unitary $\mathcal{T}$ operator can always be expressed as $\mathcal{T}_f=\sigma_{y}K|_{\mathbf{k}\rightarrow\mathbf{-k}}$ and $\mathcal{T}_b=K|_{\mathbf{k}\rightarrow\mathbf{-k}}$ for fermions and bosons, respectively.
Here $\sigma_{x,y,z}$ are the Pauli matrices acting on the two-component wavefunctions of a single Dirac cone. $\mathcal{T}$ flips the sign of momentum~($\mathbf{k}$) and $K$ is the complex conjugation.
Let us consider the surface~(say the $xy$ plane) of a 3D system with a bulk gap.
In the presence of $\mathcal{T}$, a single Dirac cone can only appear at a $\mathcal{T}$-invariant momentum, in the vicinity of which the two-band Dirac Hamiltonian is denoted by $H_{SD}(\mathbf{k})=k_x\sigma_1+k_y\sigma_2$.
Here $k_{x,y}$ are the two surface momenta and $\sigma_{1,2}$ are two linearly independent Pauli matrices. 
$\mathcal{T}$-invariance implies $\mathcal{T}$ and $H_{SD}(\mathbf{k})$ commute, requiring the existence of at least two Pauli matrices that anti-commute with $\mathcal{T}$.
For fermions, this is satisfied since all three Pauli matrices anti-commute with $\mathcal{T}_f$.
The anti-commutation relations forbid any mass term~($\sigma_3$) in $H_{SD}(\mathbf{k})$, justifying the $\mathcal{T}$-protected single Dirac surface states found in topological insulators.
For bosons, however, $\sigma_y$ is the only Pauli matrix anti-commuting with $\mathcal{T}_b$.
So $\mathcal{T}_b$ is not compatible with $H_{SD}$.
Hence $\mathcal{T}$ has to be broken to linearly split the two bands in all surface directions away from the Dirac point, in order to form a single surface Dirac cone in photonic crystals.

\subsection{Ferrimagnetic materials}

\begin{figure}[!h]
\centering
\includegraphics[width=0.5\textwidth]{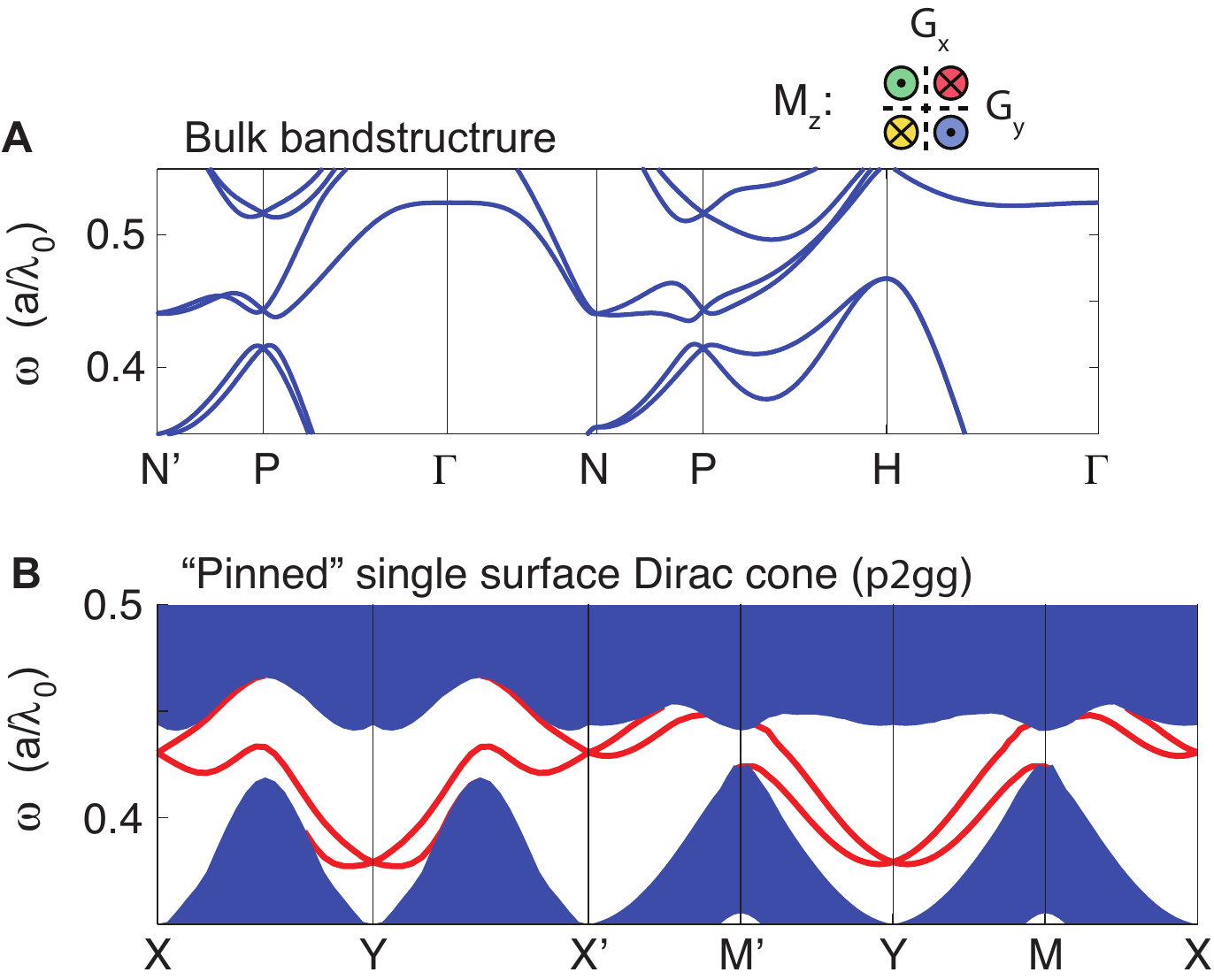}
\caption{
A) Bulk band structure showing the gap openning of the GDP.
B) Gapless $Z_2$ surface states of single Dirac cones at $X{\equiv}X'$ and $Y{\equiv}Y'$.
A perfect metallic boundary is placed from top, at the center of the cubic cell, on the (001) surface with an air gap spacing of 0.4$a$ from the photonic crystal surface.
}
\label{Fig:A}
\end{figure}

Here we break $\mathcal{T}$ with the gyromagnetic material instead of the gyroelectric material in the main text.
For example, yttrium iron garnet~(YIG) has strong gyromagnetic responses at microwave frequencies.
The permittivity and permeability of the YIG crystal can be  $\epsilon=11$,
$\mu=\left(
\begin{array}{ccc}
\mu_{\dslash} & \nu & 0\\
-\nu & \mu_{\dslash} &  0\\
0 &0 & \mu_0\\
\end{array}
\right)$, where $\nu$ is a non-zero imaginary number when the magnetization~($M_z$) is present and $\mu_0=1$.
In Fig. \ref{Fig:A}, $\mu_{\dslash}=1.5$ and $\nu=-1.2i,-1.2i,+1.2i,+1.2i$ for the red, yellow, blue and green rods respectively.
The bulk bandgap opens and the surface state has the same nontrivial connectivity as that is shown in Fig. \ref{Fig:2}C.
These calculations were performed using a modified version of the MIT Photonic Bands~\cite{Johnson2001:mpb}.

\subsection{Discussions of hybrid Wannier-center spectra}

\begin{figure}[ht]
\centering
\includegraphics[width=0.4\textwidth]{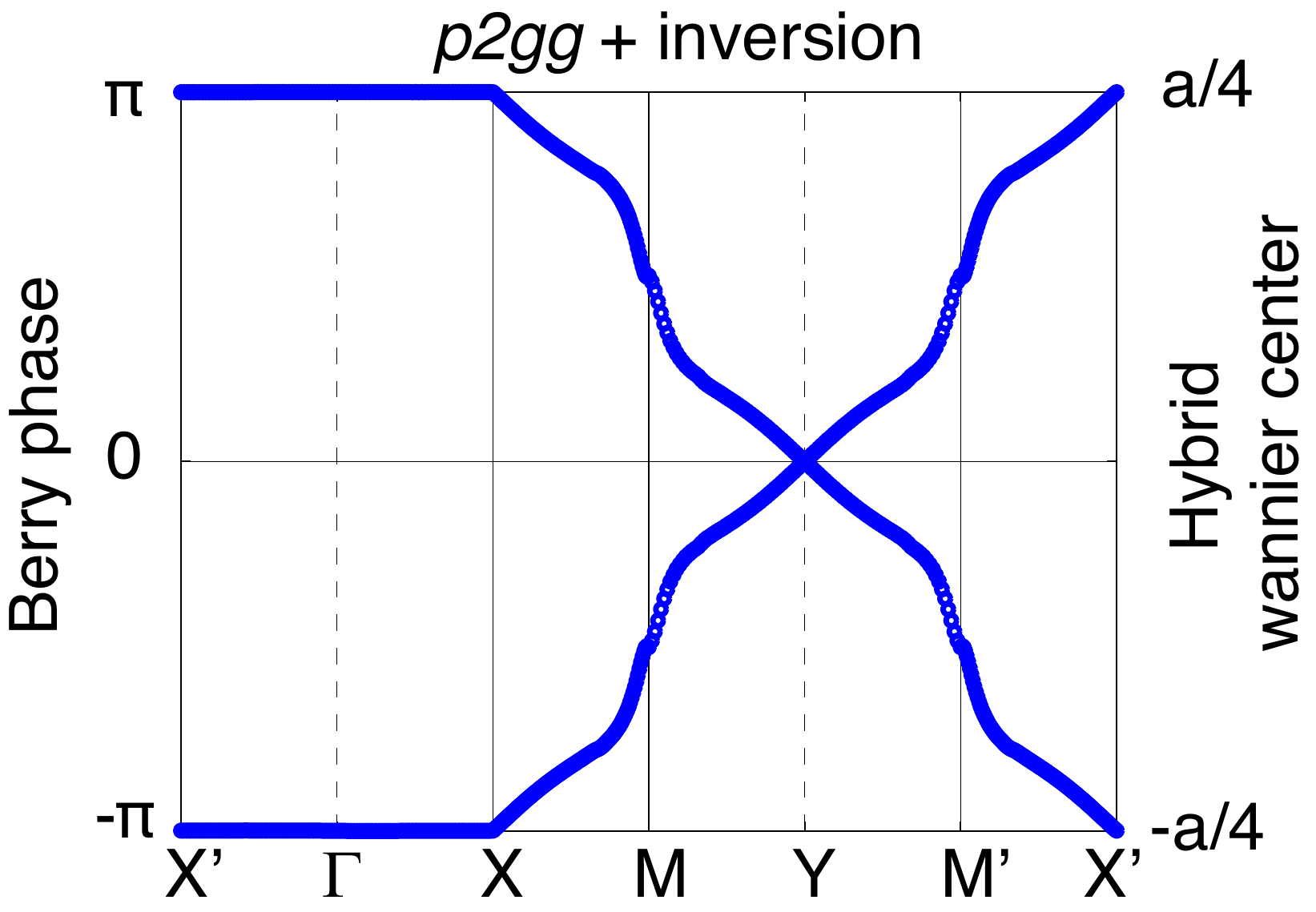}
\caption{The gapless hybrid Wannier centers corresponds to the non-trivial surface states~($p2gg$) in Fig. \ref{Fig:2}C.}
\label{Fig:B}
\end{figure}

Using the notation in Ref. \cite{taherinejad2014wannier}, the hybrid Wannier centers $\bar{z}_n(k_{\perp})$ and the equivalent Berry phases $\phi_n(k_{\perp})$ is related by $\bar{z}_n(k_{\perp})=\frac{c}{2\pi}\phi_n(k_{\perp})$. Here $n$ labels the band, $k_{\perp}$ means the wavevector in the surface BZ perpendicular to $k_z$. $c$ is the real-space period in the direction of the surface, which is $a/2$ in our bcc lattice for (001) surface.

Hybrid Wannier centers~[$\bar{z}_n(k_{\perp})$] are properties of the bulk. In addition to the symmetries on the surface, $\bar{z}_n(k_{\perp})$ also have $z$ to $-z$ symmetries in the bulk. In our system, this $z$-symmetry is inversion. so $\bar{z}_n(k_{\perp})=-\bar{z}_n(-k_{\perp})$ in all the plots in Fig. \ref{Fig:3} and \ref{Fig:B}. At inversion-invariant momenta~($X$,$M$,$\bar{\Gamma}$, $Y$) where $k=-k$ up to a reciprocal vector, $\sum\limits_{n}\bar{z}_n(k_{\perp})=0$ or $\pi$. If the Wannier centers have a single crossing in the closed loop of $X'-\bar{\Gamma}-X$ or $M-Y-M'$, the crossing point must locate at one of the inversion-invariant points.
The above arguments are all consistent with the three plots of Fig. \ref{Fig:3}A,B and Fig. \ref{Fig:B}.

Fig. \ref{Fig:B} plots the hybrid Wannier centers in the surface BZ of surface symmetry $p2gg$. Due to the high symmetry, the two Wannier centers are completely degenerate on the $X'-\bar\Gamma-X$ line at the phase value of $\pi$. The 2D vertical plane in the bulk BZ containing $X'-\bar\Gamma-X$ is the only plane on which every $k$ point is invariant under $G_y$.
Consequently, the lowest two bulk bands on this vertical plane can be separately labeled by the two $G_y$ eigenvalues of $g_y^{\pm}$.
On the other hand, the multiplication of $G_x$ and inversion~($P$) also transforms the Bloch states, within this vertical plane, from $(k_x, 0, k_z)$ into $(k_x,0,-k_z)$.
In addition, $[G_x{P},G_y]=0$.
This commutation relation means that these two operators share the same eigenstates on the plane, so that we can transforming the states within each separate branch of the two bulk bands labeled by $g_y^{\pm}$.
Since $G_x{P}$ transforms $k_z$ into $-k_z$ in the plane.
This $z$-symmetry requires $\bar{z}(\mathbf{k_{\perp}})=-\bar{z}(\mathbf{k}_{\perp})$, i.e. Berry phases of $0$ or $\pi$ for both branches of the two bulk bands, for $\mathbf{k_{\perp}}$ on the $X'-\bar{\Gamma}-X$ line.

\subsection{Calculation of Berry phases (hybrid Wannier centers)}
The multi-band non-Abelian Berry phases are calculated through the linking matrices $M_{mn}^{\mathbf{k},\mathbf{k}+\Delta{\mathbf{k}}}$ of the Bloch wavefunctions $u_{m\mathbf{k}}$ between the neighbouring two $\mathbf{k}$ points. $M_{mn}^{\mathbf{k},\mathbf{k}+\Delta{\mathbf{k}}}=\langle{u}_{m\mathbf{k}} | u_{n(\mathbf{k}+\Delta{\mathbf{k}})}\rangle$. We multiply the linking matrices to be the Wilson loop $W(\mathbf{k}_{\perp})=\prod{M^{\mathbf{k}_{\dslash},\mathbf{k}_{\dslash}+\Delta{\mathbf{k}_{\dslash}}}}$ along the closed loop in the BZ-- a parallel transport cycle. The Wilson loop eigenvalues are $\lambda_n(\mathbf{k}_{\perp})$ and the Berry phases are $\phi_n(\mathbf{k}_{\perp})=\textrm{Im}[\textrm{log}{\lambda_n({\mathbf{k}_{\perp}}})]$.

The key of this calculation is fixing the periodic gauge at the two end points~($\mathbf{k}_0$ and $\mathbf{k}_{\textrm{last}}$) differ by a reciprocal vector $\mathbf{G}$ in the bulk BZ.
We set $u(\mathbf{k}_{\textrm{last}})=u(\mathbf{k}_0+\mathbf{G})=\textrm{e}^{-i\mathbf{G}\cdot\mathbf{r}}u(\mathbf{k}_0)$.
For other $\mathbf{k}$ points in the loop, gauge fixing is not required.


\end{document}